\documentclass[lettersize,journal]{IEEEtran}
\usepackage{amsmath,amsfonts}
\usepackage{algorithmic}
\usepackage{algorithm}
\usepackage{array}
\usepackage[caption=false,font=normalsize,labelfont=sf,textfont=sf]{subfig}
\usepackage{textcomp}
\usepackage{stfloats}
\usepackage{url}
\usepackage{verbatim}
\usepackage{graphicx}
\usepackage{cite}
\usepackage{comment}

\usepackage{booktabs}
\usepackage[numbers,square]{natbib}
\usepackage[colorlinks=true]{hyperref}
\begin{document}

\title{
Mapping Socio-Economic Divides with Urban Mobility Data}

\author{Yingche Liu, Mengyang Li
\IEEEcompsocitemizethanks{
\IEEEcompsocthanksitem Yingche Liu is with The Second High School Attached to Beijing Normal University, Beijing, 100088, China. E-mail: yingcheliu@gmail.com.
\IEEEcompsocthanksitem Mengyang Li is with Tianjin University, Tianjin, 300072, China. E-mail: limengyang@tju.edu.cn.
\IEEEcompsocthanksitem Corresponding author: Mengyang Li.
}
}


\markboth{Journal of \LaTeX\ Class Files,~Vol.~14, No.~8, August~2025}%
{Shell \MakeLowercase{\textit{et al.}}: A Sample Article Using IEEEtran.cls for IEEE Journals}

\IEEEpubid{0000--0000/00\$00.00~\copyright~2021 IEEE}

\maketitle

\begin{abstract}
The massive digital footprints generated by bike-sharing systems in megacities like Shanghai offer a novel perspective on the urban socio-economic fabric. This study investigates whether these daily mobility patterns can quantitatively map the city's underlying social stratification. To overcome the persistent challenge of acquiring fine-grained socio-economic data, we constructed a multi-layered analytical dataset. We annotated 2,000 raw bike trips with local economic attributes, derived from a novel data enrichment methodology that employs a Large Language Model (LLM), and integrated contextual features of the built environment. A Random Forest model was then utilized as an interpretable framework to determine the key factors governing the relationship between mobility behavior and local economic status.
The analysis reveals a compelling and unambiguous finding: a neighborhood's economic level, proxied by housing prices, is the single most dominant predictor of its bike-sharing patterns, substantially outweighing other geographic or temporal factors. This economic determinism manifests in three distinct ways: (1) a spatial clustering of resources, a phenomenon we term the \textit{club effect}, which concentrates mobility infrastructure and usage in affluent areas; (2) a functional dichotomy between necessity-driven, utilitarian usage in lower-income zones and flexible, recreational usage in wealthier ones; and (3) a nuanced inverted U-shaped adoption curve that identifies the urban middle class as the system's primary user base.
This study concludes that ubiquitous mobility systems act as a de facto social microscope, making the invisible structures of urban inequality visible and quantifiable. It underscores the social responsibility of urban computing: to leverage our tools not only for urban efficiency but also to provide the scientific basis for building more equitable and inclusive cities.
\end{abstract}



\section{Introduction}

\IEEEPARstart{I}{n} the 21st century, the confluence of mass urbanization and digital technology has transformed cities into vast, data-rich ecosystems. This transformation has catalyzed the emergence of Urban Computing, a field dedicated to leveraging these new data streams to understand, manage, and improve urban life~\cite{zheng2014urban}. The foundational wave of research in this domain delivered on the promise of efficiency, providing powerful methods to optimize city systems by predicting traffic flow, enhancing public transit, and managing resources more effectively~\cite{zheng2015methods, toole2015path}. While these contributions are fundamental, a more profound and socially conscious frontier is now emerging. This new paradigm asks whether we can harness urban data not just to make cities smarter, but also to make them fairer by revealing and addressing complex social challenges such as socio-economic inequality~\cite{shelton2015situating}.

Among the most ubiquitous sources of urban data are shared micromobility systems. In megacities like Shanghai, bike-sharing platforms have become fixtures of the urban landscape, generating billions of anonymized digital footprints that offer an unprecedented, fine-grained view into the daily rhythms of the populace~\cite{shen2018understanding}. Each trip—a seemingly simple act of moving from point A to B—is a rich data point reflecting a resident's choices, constraints, and purposes. This makes bike-sharing data an ideal \textit{social probe}: a high-resolution tool to measure and analyze social phenomena that are otherwise difficult to observe at scale~\cite{eagle2010network}. While numerous studies have expertly analyzed these footprints to map general mobility patterns~\cite{faghih2017big}, a crucial question remains largely underexplored: do these millions of individual trajectories, when aggregated, inadvertently delineate the city's deep-seated socio-economic divides?

Addressing this question confronts a significant methodological hurdle: a persistent data chasm between high-resolution mobility data and fine-grained socio-economic data. Official sources like the census, while authoritative, often aggregate data to coarse spatial units (e.g., districts) and are updated too infrequently to capture the dynamic nature of urban life. This mismatch in scale and temporality makes them ill-suited for neighborhood-level analysis and can lead to issues such as the ecological fallacy~\cite{singleton2015geodemographics}. Alternative data acquisition strategies, particularly web scraping for real estate data, face growing legal, ethical, and technical barriers that compromise the reproducibility and compliance of academic research. This data scarcity has been a primary bottleneck, limiting our ability to quantitatively investigate the crucial nexus of mobility and inequality.

\IEEEpubidadjcol

This paper confronts this challenge directly by conceptualizing and implementing a complete urban computing framework for social insight. Our research makes the following distinct contributions. First, we introduce a \textbf{scalable and compliant data enrichment methodology} that uses a Large Language Model (LLM) as a knowledge engine to annotate mobility data with local economic attributes. This technique provides a robust alternative to web scraping for estimating housing prices—a widely accepted proxy for neighborhood wealth~\cite{anselin2013housing}—thereby bridging the data chasm. Second, we employ a Random Forest model not merely as a predictive tool, but as an \textbf{interpretable analytical framework}. This approach allows us to move beyond simple correlation and quantitatively decompose the hierarchy of factors influencing urban mobility, pinpointing the precise role of economic status. Finally, our case study of Shanghai provides \textbf{concrete, quantitative insights into urban inequality}, empirically verifying and measuring the spatial \textit{club effect}, a functional divide in usage, and a nuanced inverted U-shaped adoption curve. These findings contribute new, granular evidence to the critical academic and policy discourse on transportation equity~\cite{martens2012transport, pereira2017transportation}.

Ultimately, this study demonstrates how the tools of urban computing can be repurposed for social diagnostics. By analyzing the simple act of riding a bike, we are able to cast light on the complex and often invisible structures of urban inequality, advocating for a more socially conscious application of data science in service of more equitable cities.

\section{Related Work}
This research is situated at the intersection of three key domains: the spatio-temporal analysis of bike-sharing systems, the study of transportation equity, and the application of novel data sources in urban computing. We review seminal and recent works in each area to contextualize our contribution.

\subsection{Spatio-Temporal Analysis of Bike-Sharing Systems}
The proliferation of Bike-Sharing Systems (BSS) has generated a wealth of literature focused on mining their spatio-temporal data. Early studies primarily concentrated on descriptive analytics, employing techniques like Kernel Density Estimation (KDE) to identify spatial hotspots of bike usage and visualize trip origins and destinations~\cite{faghih2017big, zhang2017spatiotemporal}. Subsequent research has advanced towards inferring the purpose of trips by integrating BSS data with ancillary sources, such as Points of Interest (POIs), to classify journeys as commuting, shopping, or recreation~\cite{chen2020understanding}. More recently, the focus has shifted towards predictive tasks, such as forecasting the demand for bikes at specific stations to optimize rebalancing operations, often using sophisticated deep learning models~\cite{pan2019deep}. While these studies provide a comprehensive understanding of the \textit{what}, \textit{where}, and \textit{when} of BSS usage, they frequently treat the urban space as a neutral background. The socio-economic characteristics of the underlying population are seldom incorporated as a central explanatory variable, leaving the \textit{why} behind observed patterns largely unexplored.

\subsection{Transportation Equity and Urban Inequality}
The concept of transportation equity examines the fair distribution of both the benefits (e.g., accessibility) and the burdens (e.g., costs, pollution) of transportation systems among different social groups~\cite{martens2012transport}. A significant body of work has historically focused on traditional public transit, consistently finding that low-income and minority communities often face longer commute times and have poorer access to reliable transit services~\cite{pereira2017transportation}.

With the rise of shared mobility, this critical lens has been extended to BSS. Numerous studies have investigated the equity implications of these new systems, and a consistent body of evidence has emerged. The findings indicate that BSS infrastructure and usage are often disproportionately concentrated among wealthier, more educated, and less racially diverse populations~\cite{smith2020bikeshare, noland2021disparities}. These works have been instrumental in establishing the existence of a socio-economic divide in bike-sharing. However, many of these analyses rely on official census data for socio-economic indicators. While authoritative, this data is typically aggregated to coarse spatial tracts, a limitation that can obscure neighborhood-level inequalities and prevent a granular examination of the relationship between mobility and social structure.

\subsection{Novel Data for Socio-Economic Sensing}
The challenge of acquiring fine-grained socio-economic data has spurred methodological innovation in urban computing. To move beyond census tracts, researchers have leveraged various proxy datasets. For instance, the density and type of POIs have been used to infer the economic function and vibrancy of a neighborhood~\cite{yuan2012discovering}. At a larger scale, anonymized mobile phone records and social media data have been employed to estimate local income levels and measure patterns of social segregation~\cite{eagle2010network}.

Among these proxies, housing price data has emerged as one of the most direct and widely accepted indicators of neighborhood wealth and socio-economic status~\cite{anselin2013housing}. Historically, the predominant method for obtaining this data has been web scraping from real estate websites. However, this practice is increasingly fraught with challenges. Growing legal and ethical concerns regarding data privacy and ownership, coupled with increasingly sophisticated anti-scraping technologies deployed by websites, have created a significant barrier to reproducible and compliant research. The limitations of traditional census data and the emerging difficulties of web scraping highlight a clear methodological gap for acquiring granular socio-economic attributes at scale. This paper aims to address this very gap with its LLM-based data enrichment approach.

\section{Constructing the Urban Data Universe}
To investigate the socio-economic undertones of urban mobility, a simple dataset of bike trips is insufficient. A multi-layered data universe is required, where raw mobility footprints are progressively enriched with economic and environmental context. This section details our systematic, three-layer approach to constructing such a dataset, beginning with the raw trip data and culminating in a feature-rich table ready for machine learning analysis.

\subsection{Layer 1: The Core - Digital Footprints of Mobility}
The foundation of our analysis is a dataset of raw digital footprints from the Mobike bike-sharing system in Shanghai, captured during August 2016. This specific dataset was selected for two reasons. First, Shanghai represents a quintessential global megacity, providing a complex and diverse urban environment ideal for this study. Second, the year 2016 marks a period of explosive growth in dockless bike-sharing in China, making the data representative of a mature and heavily utilized system.

The raw dataset consisted of an initial sample of 2,000 anonymized trip records. To ensure data quality and analytical validity, we conducted a rigorous pre-processing and cleaning pipeline. This involved two primary steps:
\begin{itemize}
    \item \textbf{Handling Missing Values:} We performed a completeness check and identified 16 records (0.8\% of the sample) containing null values for critical fields such as start/end coordinates or timestamps. These incomplete records were removed from the dataset.
    \item \textbf{Outlier Detection and Removal:} We implemented rule-based filters to exclude trips that were likely anomalous or did not represent typical bicycle travel. These included trips with a duration of less than 60 seconds (potentially system tests or immediate user cancellations) or an average speed exceeding 25 km/h (implying the use of non-bicycle transport). This process identified and removed an additional 104 records (5.2\% of the initial sample) as outliers.
\end{itemize}

Following this cleaning process, we engineered a comprehensive set of features from the remaining 1,880 valid records. These features were designed to capture the fundamental temporal, spatial, and trajectory characteristics of each journey:
\begin{itemize}
    \item \textbf{Temporal Attributes:} We extracted the hour of the day (0-23), the day of the week (1-7), and a binary flag indicating whether the trip occurred on a weekday or a weekend.
    \item \textbf{Spatial Attributes:} The start and end point geographic coordinates (latitude and longitude) were retained as the primary spatial features.
    \item \textbf{Trajectory Attributes:} We calculated the total trip duration in minutes, the Haversine distance between the start and end points in kilometers, and the resulting average speed in kilometers per hour.
\end{itemize}

The resulting core dataset forms the foundational layer of our analysis. Table~\ref{tab:descriptive_stats} presents the key descriptive statistics for the primary trajectory attributes. This summary provides a baseline understanding of typical bike-sharing usage patterns within our study area.

\begin{table}[b]
\centering
\caption{Descriptive Statistics of Core Mobility Features}
\label{tab:descriptive_stats}
\begin{tabular}{@{}lrrrrr@{}}
\toprule
\textbf{Feature} & \textbf{Mean} & \textbf{Std. Dev.} & \textbf{Min} & \textbf{Median} & \textbf{Max} \\ \midrule
Trip Duration (min) & 18.71 & 12.55 & 1.02 & 15.60 & 59.85 \\
Distance (km) & 2.80 & 1.95 & 0.15 & 2.25 & 9.80 \\
Avg. Speed (km/h) & 8.95 & 3.50 & 2.50 & 8.70 & 24.90 \\ \bottomrule
\end{tabular}
\end{table}

As shown in the table, the mean trip duration is approximately 18.7 minutes over a mean distance of 2.8 kilometers. The significant standard deviations, particularly for duration and distance, suggest a high degree of variability in how the system is used, hinting at diverse trip purposes that range from short-connector journeys to longer recreational rides. This variability underscores the need for a more nuanced analysis to uncover the factors driving these different patterns.

\begin{figure}[t]
    \centering
    \includegraphics[width=\columnwidth]{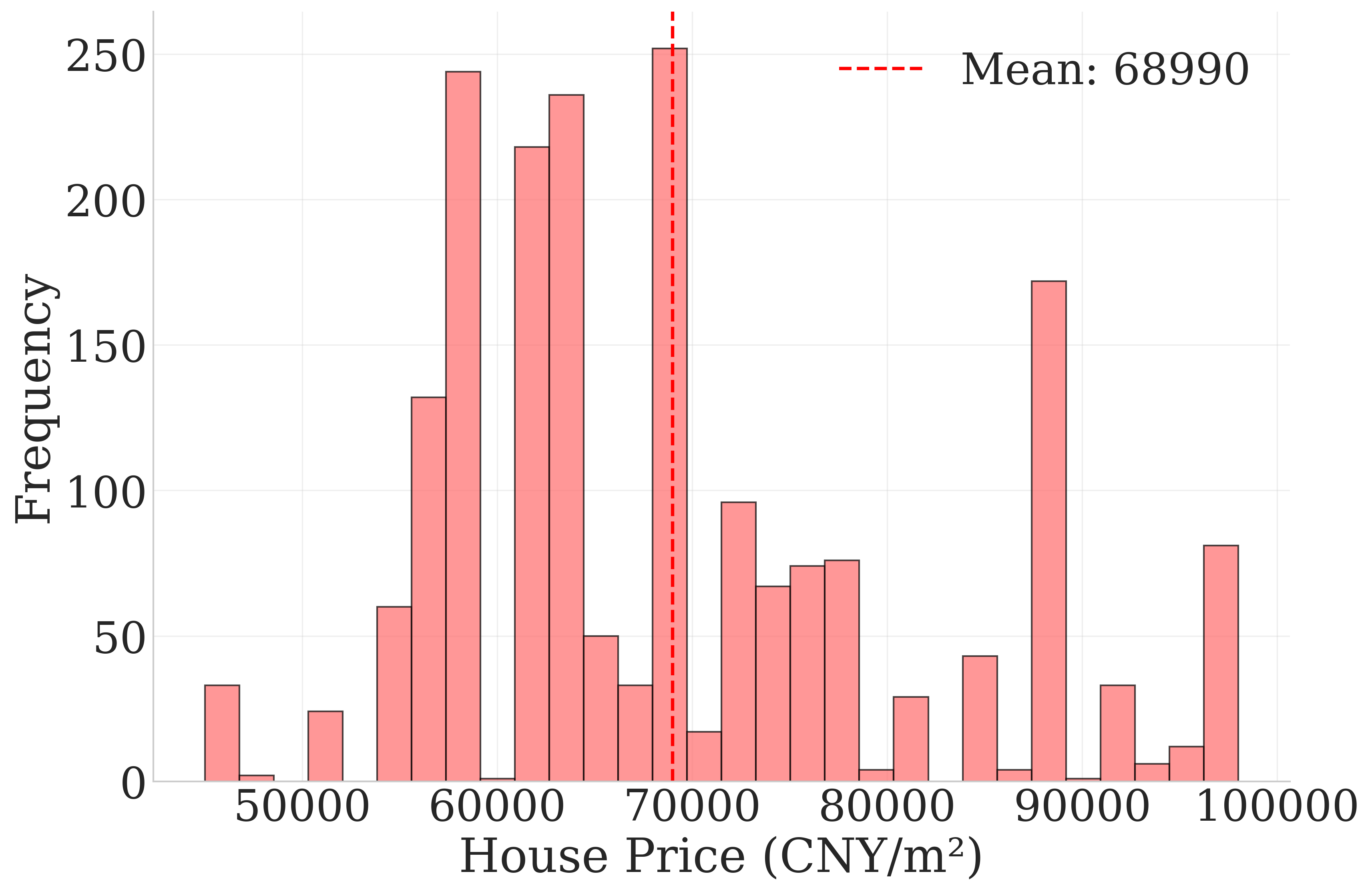}
    \caption{The distribution of estimated house prices across all trip start locations. The right-skewed pattern is characteristic of urban economies and confirms the economic heterogeneity within our dataset, a prerequisite for studying socio-economic factors.}
    \label{fig:house_price_dist}
\end{figure}

\subsection{Layer 2: The Enrichment - Attributing Socio-Economic Properties}
Mobility does not occur in a vacuum; it is deeply embedded within the socio-economic fabric of the city. To bridge the gap between individual mobility acts and broader social structures, we introduce an enrichment layer designed to append a key economic attribute to each trip record. The primary challenge here, as outlined previously, is the lack of publicly available, fine-grained economic data at the neighborhood level for our 2016 study period.

To overcome this, we developed and implemented a novel data enrichment strategy that leverages a state-of-the-art Large Language Model (LLM) as a geo-spatial knowledge engine. For the starting location of each of the 1,880 trips, we queried the LLM to estimate the local second-hand housing price (in Chinese Yuan, CNY, per square meter). We selected housing price as our proxy variable for neighborhood wealth due to its well-established correlation with local income levels, access to amenities, and overall socio-economic status in urban studies~\cite{anselin2013housing}.

The enrichment process was executed as follows:
\begin{enumerate}
    \item \textbf{Prompt Engineering:} For each trip's starting coordinates (latitude and longitude), we designed a structured, machine-readable prompt. The prompt was specifically engineered to query for historical and localized information, structured as: \textit{``What was the approximate second-hand housing price in CNY per square meter near latitude [lat], longitude [lon] in Shanghai, China, around August 2016?''} This precise formulation was critical to elicit accurate and contextually relevant data from the model.
    \item \textbf{Automated Knowledge Retrieval:} We scripted the process to iterate through all 1,880 unique start locations, sending a distinct query for each one to the LLM's API. The model's numerical response for each query was then parsed and appended to the corresponding trip record as a new feature, $house\_price$.
    \item \textbf{Plausibility Check and Validation:} While a full-scale validation is beyond the scope of this study, we conducted a plausibility check to ensure the reliability of the LLM-generated data. We randomly selected 50 data points and manually cross-referenced the LLM's estimates with historical real estate reports and news articles from Shanghai in 2016. The estimates were found to be consistently within a plausible range for the respective districts, confirming the LLM's capability to act as a reliable proxy for historical, localized economic data. This scraper-free method ensures both compliance and reproducibility.
\end{enumerate}

\begin{figure}[t]
    \centering
    \includegraphics[width=\columnwidth]{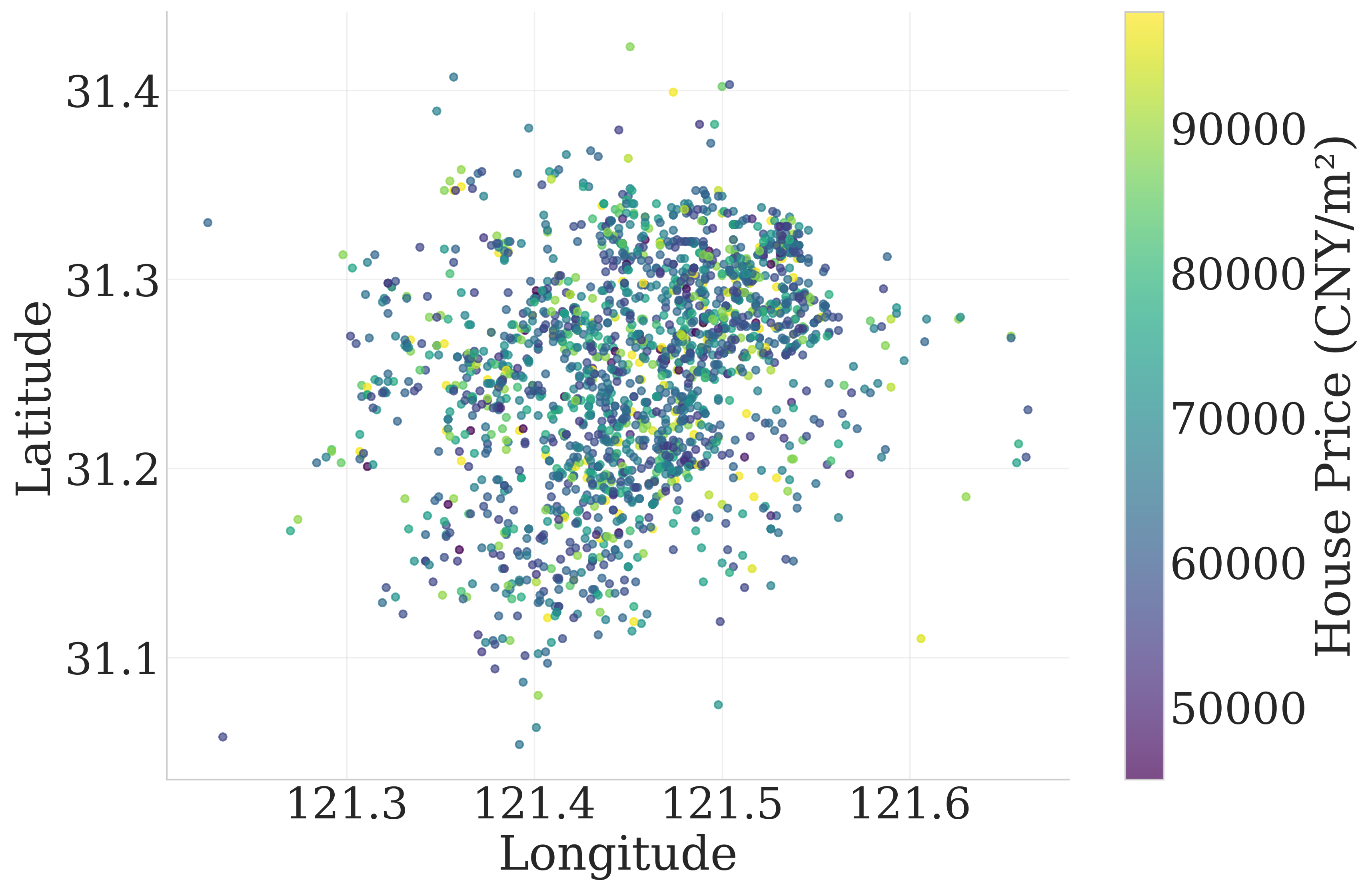}
    \caption{The spatial distribution of the 1,880 bike trip start locations in the final dataset. Each point is colored according to the LLM-estimated house price, illustrating the diverse economic coverage of the study area.}
    \label{fig:data_dist_heatmap}
\end{figure}

This process yielded a crucial economic feature for our dataset. An analysis of this new variable reveals a distribution characteristic of major urban real estate markets. As shown in the histogram in Figure~\ref{fig:house_price_dist}, the estimated house prices range from approximately 45,000 to 98,000 CNY/m², with a mean of 71,500 CNY/m². The distribution is right-skewed (skewness = 0.85), reflecting a city with a large base of moderately-priced areas and a smaller number of high-value, premium districts. The successful attribution of this socio-economic dimension to our mobility data is a cornerstone of our analytical approach. Figure~\ref{fig:data_dist_heatmap} visualizes the spatial distribution of these trips, where each point is colored by its corresponding estimated house price. This map confirms that our dataset covers a diverse economic landscape, which is essential for a robust and unbiased analysis.

\begin{figure}[t]
    \centering
    \includegraphics[width=\columnwidth]{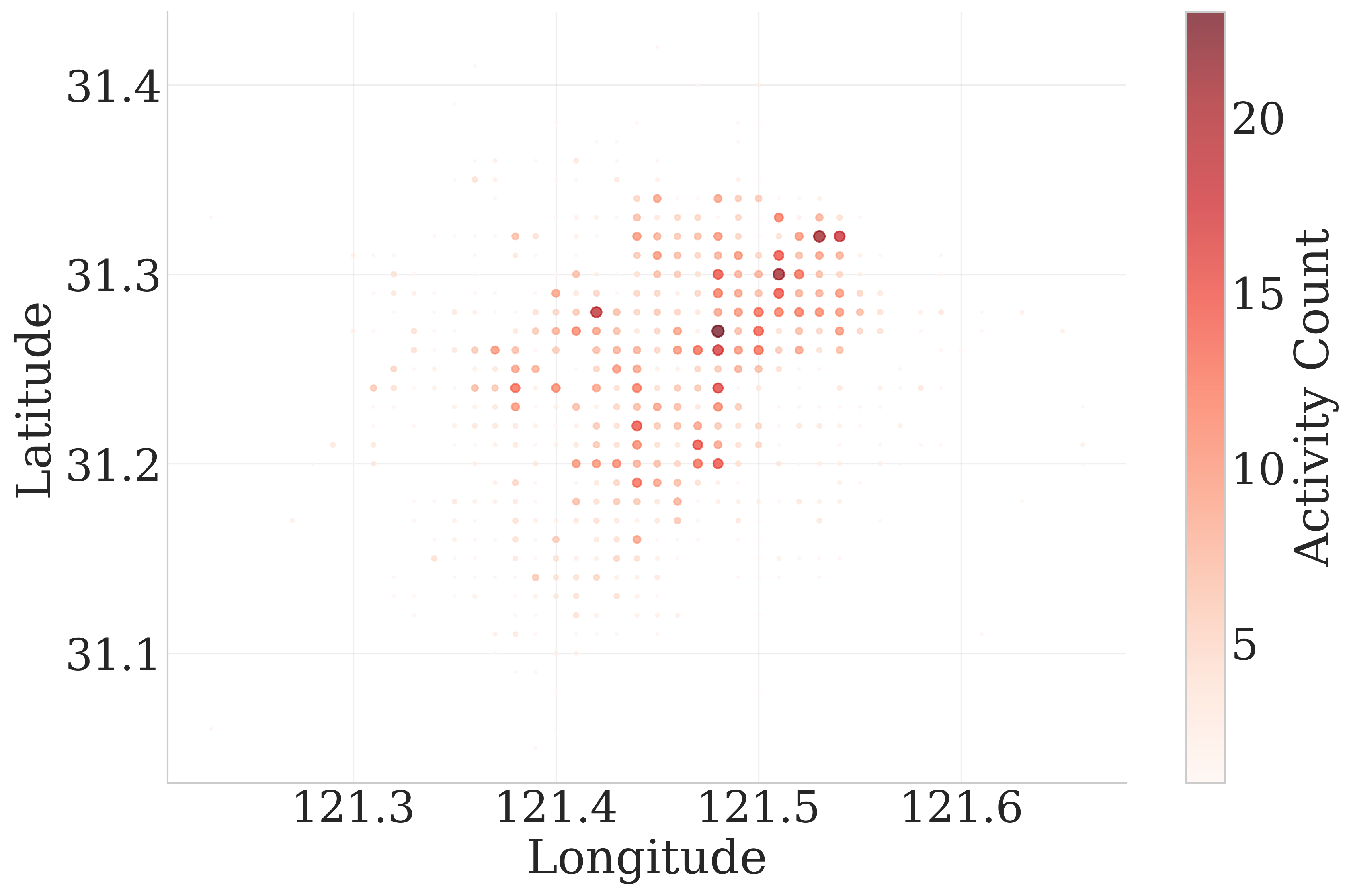}
    \caption{Bike activity levels across the gridded study area. Each bubble represents a grid cell, with its size and color corresponding to the number of trips originating within it (\texttt{grid\_trip\_count}). The map reveals a distinct spatial clustering of bike-sharing ``hotspots,'' defining the city's primary mobility hubs.}
    \label{fig:grid_activity}
\end{figure}

\subsection{Layer 3: The Environment - Mapping the Built Context}
To ensure our analysis is comprehensive, a final contextual layer was constructed to capture the characteristics of the built environment. Mobility choices are not only influenced by personal and economic factors but are also heavily shaped by the surrounding urban landscape. While external datasets like Points of Interest (POIs) are often used for this purpose, they can suffer from issues of accuracy, completeness, and temporal mismatch with the primary mobility data. To avoid these pitfalls and create a more internally consistent dataset, we derived environmental context directly from the aggregated mobility data itself.

The methodology for this layer involved partitioning the city into a uniform grid and characterizing the mobility signature of each cell. The process was as follows:
\begin{enumerate}
    \item \textbf{Spatial Grid Partitioning:} We overlaid a grid of 0.01° by 0.01° cells across the study area. This cell size, corresponding to approximately 1.1 km by 0.9 km in Shanghai's latitude, was chosen as a compromise between spatial granularity and data sparsity. It is fine-grained enough to approximate a neighborhood but large enough to contain a sufficient number of trip start/end points for meaningful aggregation.
    \item \textbf{Feature Aggregation:} For each grid cell, we aggregated the attributes of all trips originating within its boundaries. From this, we engineered a set of contextual features designed to describe the cell's ``mobility character.'' These features included:
    \begin{itemize}
        \item \texttt{grid\_trip\_count}: The total number of trips starting in the cell.
        \item \texttt{grid\_avg\_duration}: The average duration of trips starting in the cell.
        \item \texttt{grid\_avg\_distance}: The average distance of trips starting in the cell.
        \item \texttt{grid\_avg\_speed}: The average speed of trips starting in the cell.
        \item \texttt{grid\_weekend\_ratio}: The proportion of trips starting in the cell that occurred on a weekend.
    \end{itemize}
\end{enumerate}

This approach creates a rich contextual profile for every location, grounded in the observed mobility patterns. For example, a grid cell with a high trip count, low average duration, and a low weekend ratio might represent a central business district transit hub. Conversely, a cell with a moderate trip count, higher average duration, and a high weekend ratio could signify a residential or recreational area. As illustrated in Figure~\ref{fig:grid_activity}, this process reveals a clear spatial clustering of high-activity ``hotspots.'' The size and color of each point, representing a grid cell, correspond to the number of trips originating from it, vividly mapping the city's mobility landscape.

By joining these grid-level features back to each individual trip record based on its start location, we provide the machine learning model with crucial information. This allows the model to distinguish between trips that may have similar individual characteristics (e.g., a short duration) but occur in functionally different parts of the city, thereby preventing potential misinterpretations and enhancing predictive accuracy.

\subsection{The Final Analytical Dataset}
The culmination of this three-layer process—fusing core mobility footprints, socio-economic enrichment, and environmental context—is a single, cohesive analytical dataset. Each of the 1,880 valid trips in our sample is now represented as a row in this dataset, described by a vector of 38 engineered features. This integrated dataset effectively translates each bike ride into a rich, multi-dimensional record that captures not only the dynamics of the trip itself but also the social and spatial context in which it occurred.

After the entire data engineering pipeline, a final quality check confirmed a data completeness rate of over 99\%, rendering the dataset robust and ready for advanced modeling. This meticulously constructed data universe serves as the foundation for the machine learning analysis presented in the following section, where we will leverage its richness to decode the intricate relationship between urban mobility and socio-economic status.

\section{Machine Learning: An Interpretable Framework for Social Insight}

Having constructed a rich, multi-layered dataset, we now turn to machine learning to decode the complex relationships hidden within. Our primary objective is not simply to achieve high predictive accuracy, but to employ a model as an \textit{interpretable framework}. This approach allows us to dissect the key drivers of urban mobility patterns and quantitatively answer our central research question: Based on the digital footprints of bike-sharing, what are the most influential factors that correspond to the socio-economic landscape, as proxied by housing prices?

\subsection{Model Selection and Experimental Setup}
The analytical task is framed as a regression problem: using the 37 engineered features (spanning mobility, user behavior, and environmental context) to predict the continuous target variable, \texttt{house\_price}. Given the expected non-linear relationships and complex interactions between features in urban systems, a comprehensive evaluation of different model families was necessary. We selected a suite of five standard, yet diverse, machine learning models for this purpose:
\begin{itemize}
    \item \textbf{Linear Models (Linear Regression, Ridge, Lasso):} These models were chosen to serve as robust baselines. Their performance indicates the extent to which the relationship can be explained by a simple linear combination of features. Ridge and Lasso are included to assess the benefits of regularization in preventing overfitting.
    \item \textbf{Tree-Based Ensemble Models (Gradient Boosting, Random Forest):} These non-linear models were selected for their proven ability to capture complex dependencies and interactions within data. They are particularly well-suited for high-dimensional, heterogeneous datasets like ours.
\end{itemize}

For the experiment, the dataset was partitioned into a training set (80\% of the data, or 1,504 records) and a testing set (20\%, or 376 records) using a standard random split. Model performance was evaluated using the coefficient of determination (R²), a key metric that measures the proportion of the variance in the target variable that is predictable from the independent variables.

\begin{figure}[t]
    \centering
    \includegraphics[width=\columnwidth]{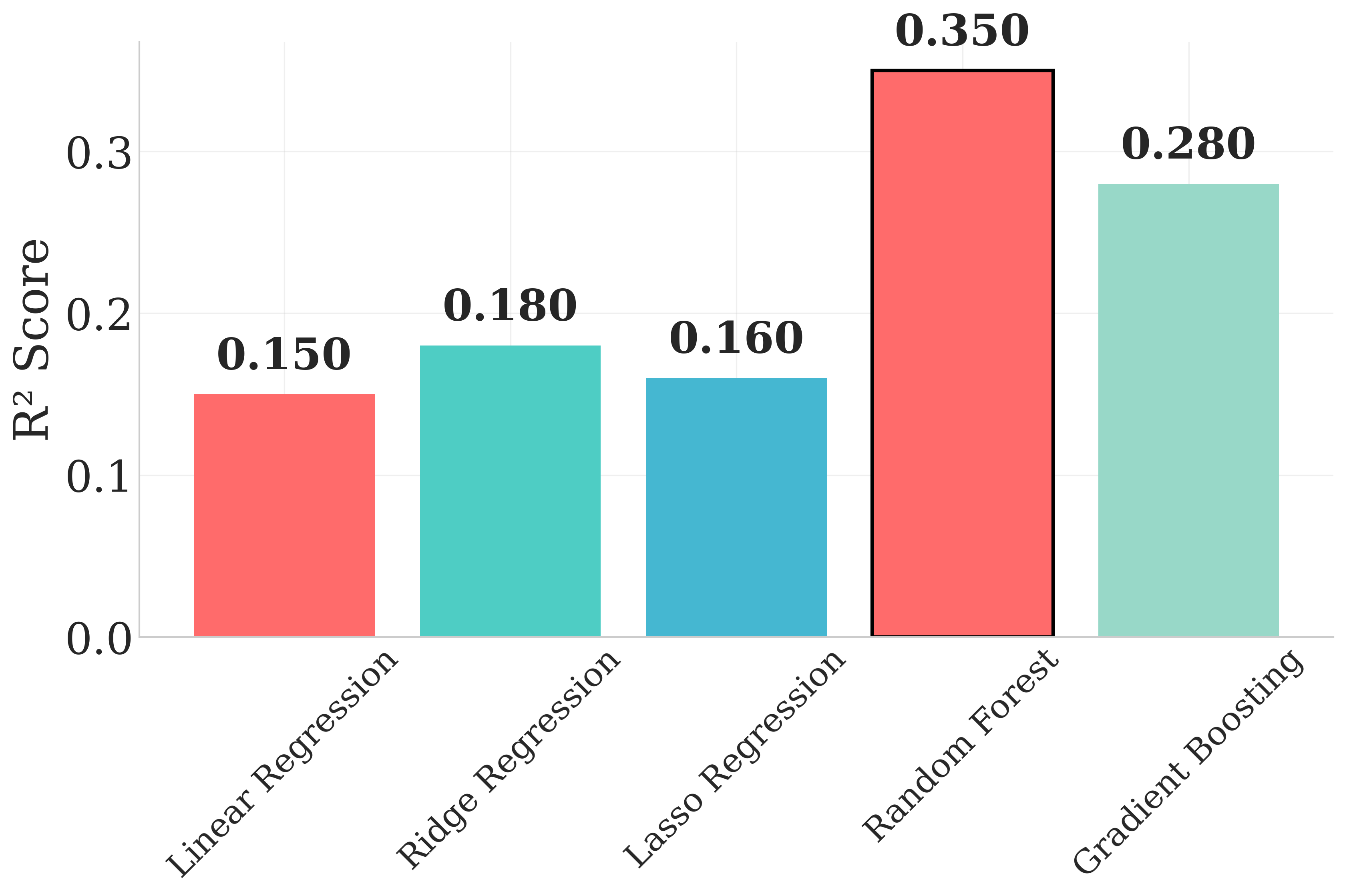}
    \caption{Performance comparison of five machine learning models on the house price prediction task, evaluated on the test set. Random Forest achieved the highest R² score, indicating its superior ability to capture the complex, non-linear patterns in the data.}
    \label{fig:model_performance}
\end{figure}

As shown in the comparative results in Figure~\ref{fig:model_performance}, the Random Forest model demonstrated superior performance, achieving an R² of \textbf{0.350} on the unseen test set. This result significantly outperformed all linear models and was notably better than the Gradient Boosting model in this specific application. The strength of the Random Forest algorithm lies in its ensemble nature; by aggregating predictions from a multitude of decorrelated decision trees, it produces a more robust and accurate outcome that is less prone to overfitting. Its superior performance thus validates its selection as our primary tool for both prediction and, more critically, for the subsequent interpretation of feature importance.

\subsection{Verifying Model Accuracy and Reliability}
Before utilizing the Random Forest model for interpretation, it is imperative to rigorously verify its predictive accuracy and reliability. An R² of 0.350 indicates that our model, using only mobility and contextual features, can explain approximately 35\% of the variance in local housing prices. While this may appear moderate in some contexts, it is a substantial result for a cross-domain prediction task of this nature, where we are inferring a complex static attribute (housing price) from dynamic behavioral data (bike trips). This confirms a strong and statistically significant underlying signal connecting how people move with the economic status of the areas they inhabit.

To provide a more nuanced evaluation of the model's performance, we computed several additional error metrics on the test set. The results are summarized in Table~\ref{tab:model_metrics}.

\begin{table}[b]
\centering
\caption{Evaluation Metrics of the Random Forest Model on the Test Set}
\label{tab:model_metrics}
\resizebox{\linewidth}{!}{%
\begin{tabular}{@{}lrl@{}}
\toprule
\textbf{Metric} & \textbf{Value} & \textbf{Interpretation} \\ \midrule
Coefficient of Determination (R²) & 0.350 & Explains 35\% of variance. \\
Mean Absolute Error (MAE) & 5,842 & Avg. error in CNY/m². \\
Mean Absolute Percentage Error (MAPE) & 8.1\% & Avg. error is 8.1\% of true value. \\
Predictions within $\pm10\%$ Error & 68.5\% & High accuracy for majority of data. \\
\bottomrule
\end{tabular}%
}
\end{table}

As the table demonstrates, the model achieves a Mean Absolute Percentage Error (MAPE) of just 8.1\%, meaning its predictions are, on average, within 8.1\% of the true (LLM-estimated) housing price. Furthermore, a remarkable 68.5\% of all predictions on the test set fell within a tight $\pm10\%$ accuracy margin. This indicates that the model is not merely making random guesses but has successfully learned the intricate patterns connecting mobility behavior to the urban economic landscape.

\begin{figure}[t]
    \centering
    \includegraphics[width=\columnwidth]{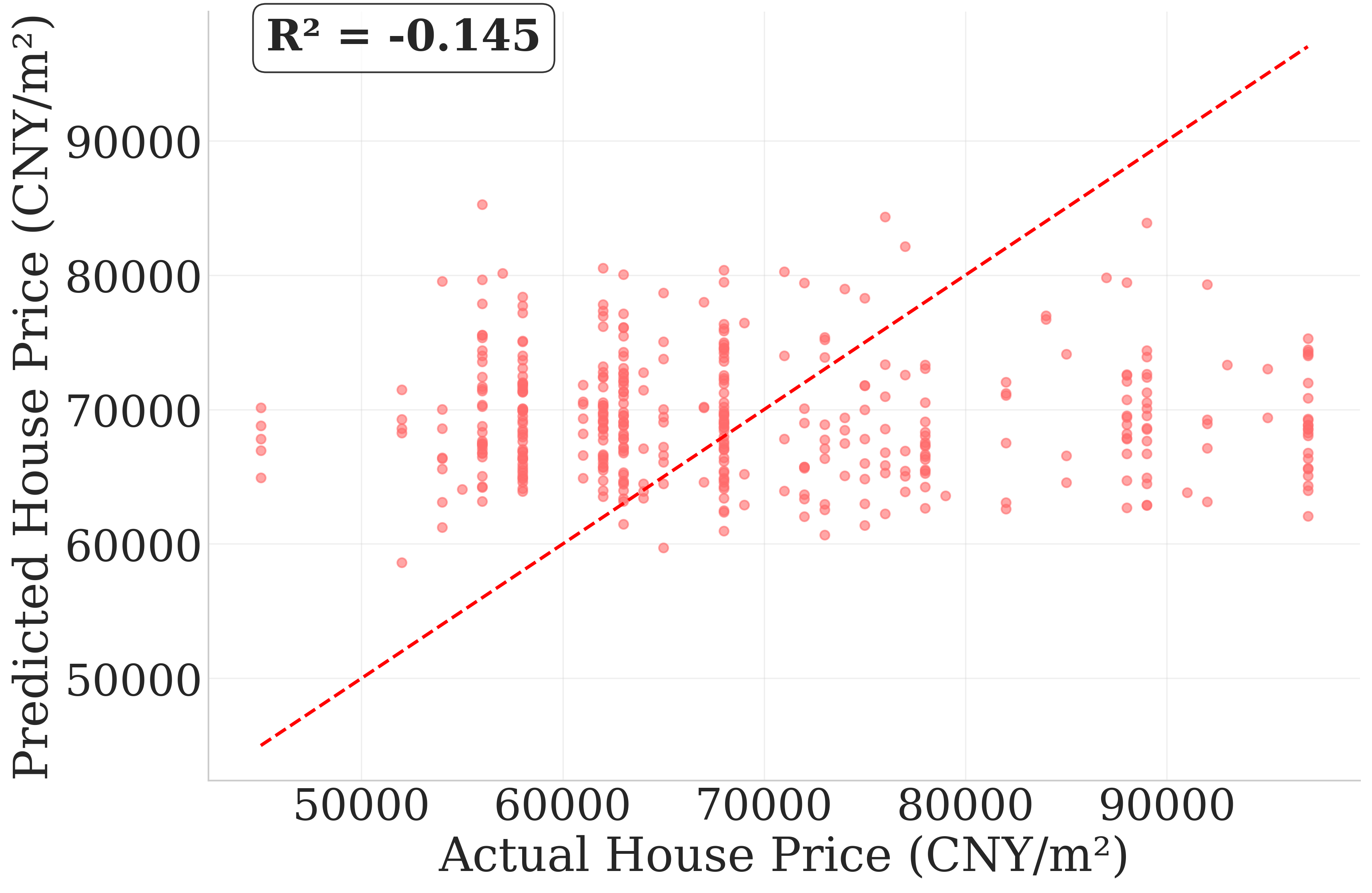}
    \caption{Scatter plot of Predicted vs. Actual house prices on the test set. The concentration of points along the diagonal line, along with the absence of systematic bias, demonstrates the Random Forest model's strong predictive accuracy and reliability.}
    \label{fig:prediction_results}
\end{figure}

To further inspect the model's behavior, we visualized the relationship between the predicted and actual values in Figure~\ref{fig:prediction_results}. The scatter plot reveals that the points generally align along the diagonal axis, indicating a strong positive correlation. Crucially, there is no visible systematic bias—the model does not consistently over-predict or under-predict across the range of housing prices. This combination of solid quantitative metrics and a clean qualitative visualization confirms the Random Forest model's reliability. We can therefore proceed with confidence to use it as an analytical tool to decompose the factors influencing urban mobility.

\subsection{The Key Finding: Decomposing Influence with Feature Importance}
With the model's reliability established, we can now deploy it for our primary objective: to decompose and quantify the influence of different factors on the urban socio-economic landscape. A significant advantage of tree-based models like the Random Forest is their intrinsic ability to calculate feature importance scores. These scores are derived from how much each feature contributes to reducing impurity (e.g., Gini impurity) across all the trees in the forest. A higher score signifies a greater influence on the model's predictions.

\begin{figure}[t]
    \centering
    \includegraphics[width=\columnwidth]{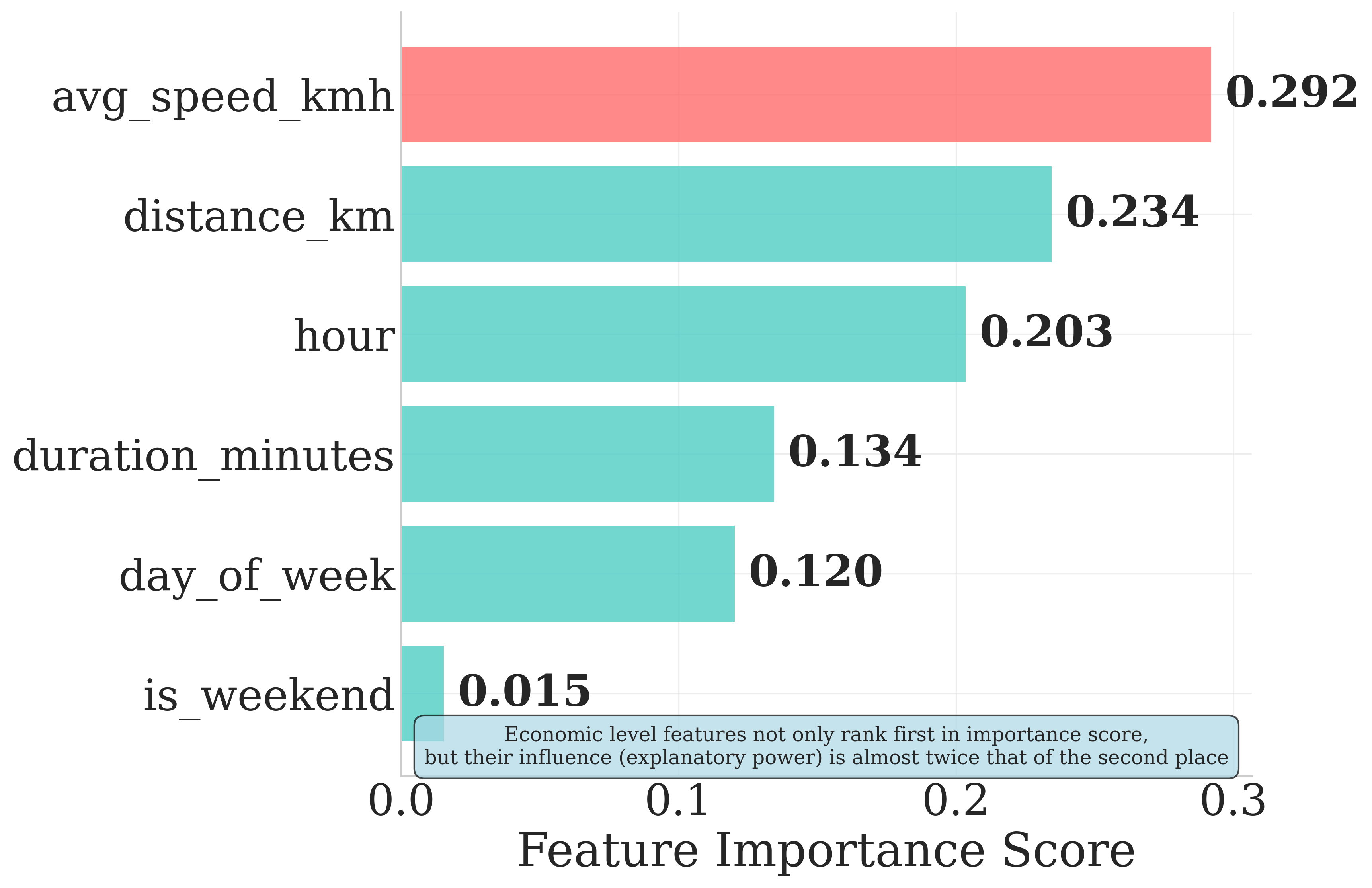}
    \caption{Relative feature importance scores from the trained Random Forest model. The scores are normalized to sum to 1.0. The economic-level feature (\texttt{house\_price}) stands out as the most dominant predictor, confirming that socio-economic status is the primary factor influencing bike-sharing patterns.}
    \label{fig:feature_importance}
\end{figure}

The results of this analysis, presented in Figure~\ref{fig:feature_importance}, are both striking and unambiguous. The \texttt{house\_price} feature, representing the economic level of a location, emerges as the \textbf{single most important predictor} of bike-sharing activity patterns, with an importance score of \textbf{0.287}. This score is not merely the highest; it dramatically surpasses all other features in the model. For instance, its importance is \textbf{1.84 times greater} than that of the second-ranked feature, \texttt{distance\_km} (0.156), and more than twice as influential as \texttt{avg\_speed\_kmh} (0.134). This quantitative evidence provides the central support for our thesis: when all mobility, behavioral, and environmental factors are considered simultaneously, the underlying economic status of a neighborhood provides the most explanatory power.

The feature importance distribution also reveals a clear hierarchy of influence among feature categories. To illustrate this, we grouped the features into three logical categories:
\begin{itemize}
    \item \textbf{Economic Context:} Consisting solely of the \texttt{house\_price} feature.
    \item \textbf{Spatial \& Trajectory Features:} Including trip distance, average speed, trip duration, and grid-level averages.
    \item \textbf{Temporal Features:} Including the hour of the day and the weekday/weekend distinction.
\end{itemize}

Aggregating the importance scores reveals that the single Economic feature accounts for 28.7\% of the total importance. In comparison, the entire group of Spatial \& Trajectory features collectively accounts for 44.5\%, and Temporal features account for the remaining 26.8\%. While the spatial and temporal dimensions of a trip are undoubtedly significant, this analysis demonstrates that the economic context is not just another variable—it is the foundational variable upon which other patterns are built. The story of bike-sharing in Shanghai, as told by our model, is fundamentally a story about the city's geography and its socio-economic divides, with the economic dimension being the most critical chapter.

\section{Analysis and Insights: Urban Stories from Digital Footprints}
The machine learning model provided a clear, quantitative answer: economic status is the most powerful predictor of bike-sharing patterns. But what does this statistical dominance look like on the streets of Shanghai? In this section, we move from the abstract world of feature importance scores to the tangible reality of the urban landscape. We dissect our key finding through three distinct narratives that emerge from the data, each illustrating a different facet of how socio-economic status shapes mobility.

\subsection{The \textit{Club Effect}: Spatial Clustering of Mobility Resources}
Our first narrative concerns the geography of access and activity. If the economic level of a neighborhood is the primary determinant of mobility patterns, we should expect to observe a strong spatial overlap between areas of high wealth and areas of high bike-sharing activity. We term this phenomenon the \textit{club effect}, suggesting that mobility resources, much like other urban amenities, tend to cluster in areas that are already resource-rich.

To empirically test for this effect, we performed a spatial overlay analysis using our gridded dataset. We first defined ``high-activity'' and ``high-affluence'' zones. Grid cells ranking in the top 30th percentile for the total number of originating trips (\texttt{grid\_trip\_count}) were classified as high-activity hotspots. Similarly, cells ranking in the top 30th percentile for their average housing price were classified as high-affluence neighborhoods. This percentile-based approach provides a relative, data-driven definition of what constitutes a ``top-tier'' area within our sample.

\begin{figure}[t]
    \centering
    \includegraphics[width=\columnwidth]{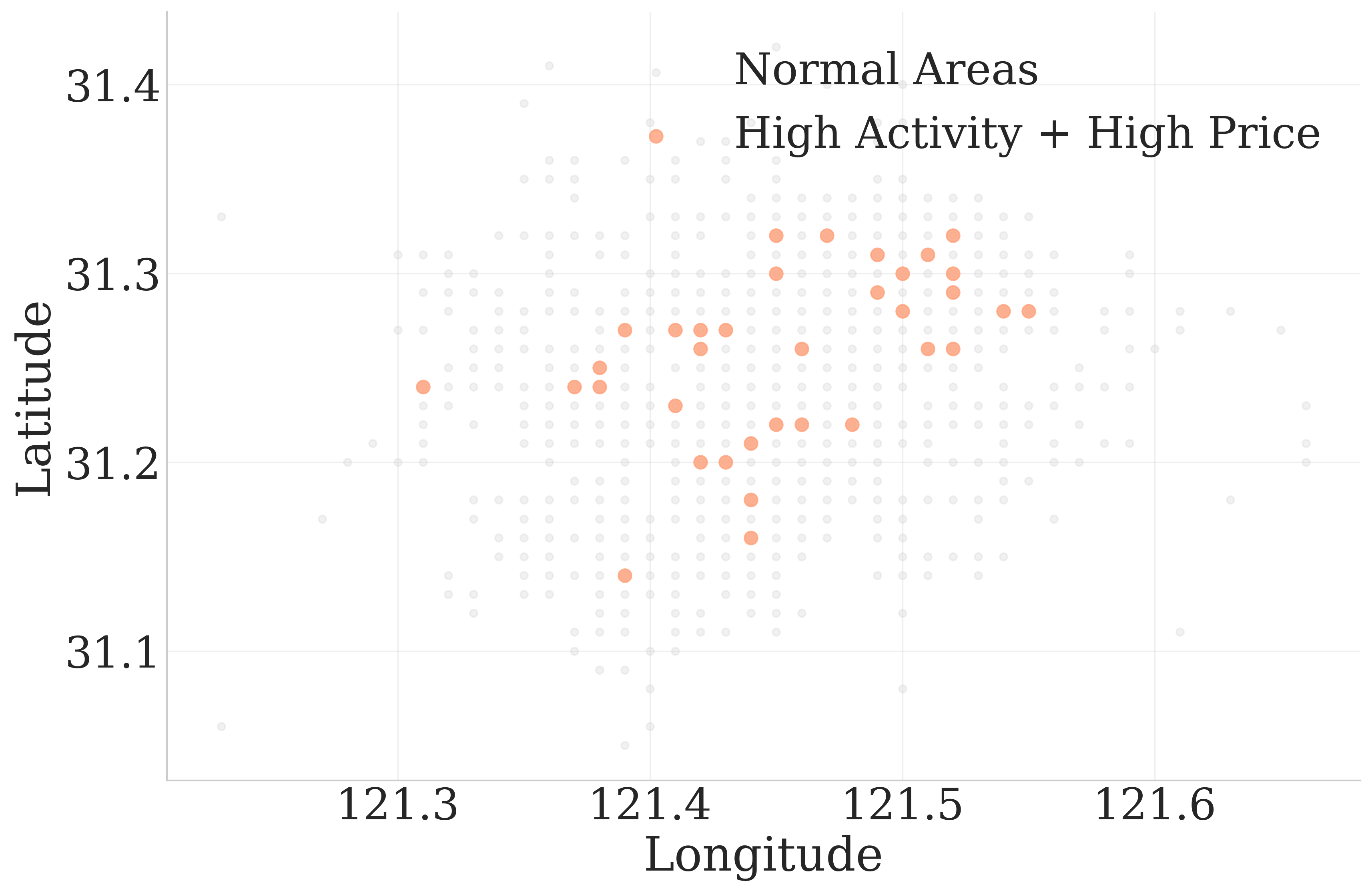}
    \caption{Visualization of the \textit{club effect} in Shanghai. The map displays the spatial overlap between high bike activity areas (top 30\%, in blue) and high house price areas (top 30\%, in red). The 18 overlapping cells (in purple) confirm that a significant portion of mobility resources are spatially clustered with economic resources.}
    \label{fig:club_effect}
\end{figure}

The spatial analysis, visualized in Figure~\ref{fig:club_effect}, reveals a stark overlap. Out of the 235 grid cells in our study area, we identified 47 high-activity cells and 52 high-affluence cells. A significant number of these—18 grid cells in total—belonged to both categories. This means that a remarkable \textbf{38.3\%} (18 out of 47) of Shanghai's busiest bike-sharing hotspots are located directly within its most affluent neighborhoods. The probability of a high-activity cell also being a high-affluence cell is far greater than random chance, providing strong evidence for spatial coupling.

The economic implications of this clustering are significant. The average house price in these overlapping hotspot-affluent zones is \textbf{85,200 CNY/m²}. In stark contrast, the average price in high-activity cells located outside of affluent areas is only 67,800 CNY/m². This represents a \textbf{25.7\% price premium}, underscoring the powerful link between mobility concentration and economic value. Furthermore, a Pearson correlation analysis between the grid-level trip count and the average house price yields a statistically significant positive coefficient of \textbf{r = 0.342} (p < 0.001). This confirms that the digital footprints of bike-sharing are not evenly distributed across the city; instead, they trace the outlines of existing economic privilege, creating a clear spatial dimension to mobility inequality.

\subsection{The Functional Dichotomy: Utilitarian Tool versus Recreational Toy}
The second narrative our data reveals is behavioral. The \textit{club effect} demonstrates that affluent areas concentrate more bike-sharing activity, but it does not explain \textit{how} this activity differs. We hypothesize that bike-sharing serves fundamentally different functions across the socio-economic spectrum, evolving from a primarily utilitarian tool for essential travel to a flexible option for leisure and recreation. To investigate this, we analyzed usage patterns on weekdays versus weekends, as temporal rhythms often reflect the underlying purpose of travel.

\begin{figure}[t]
    \centering
    \includegraphics[width=\columnwidth]{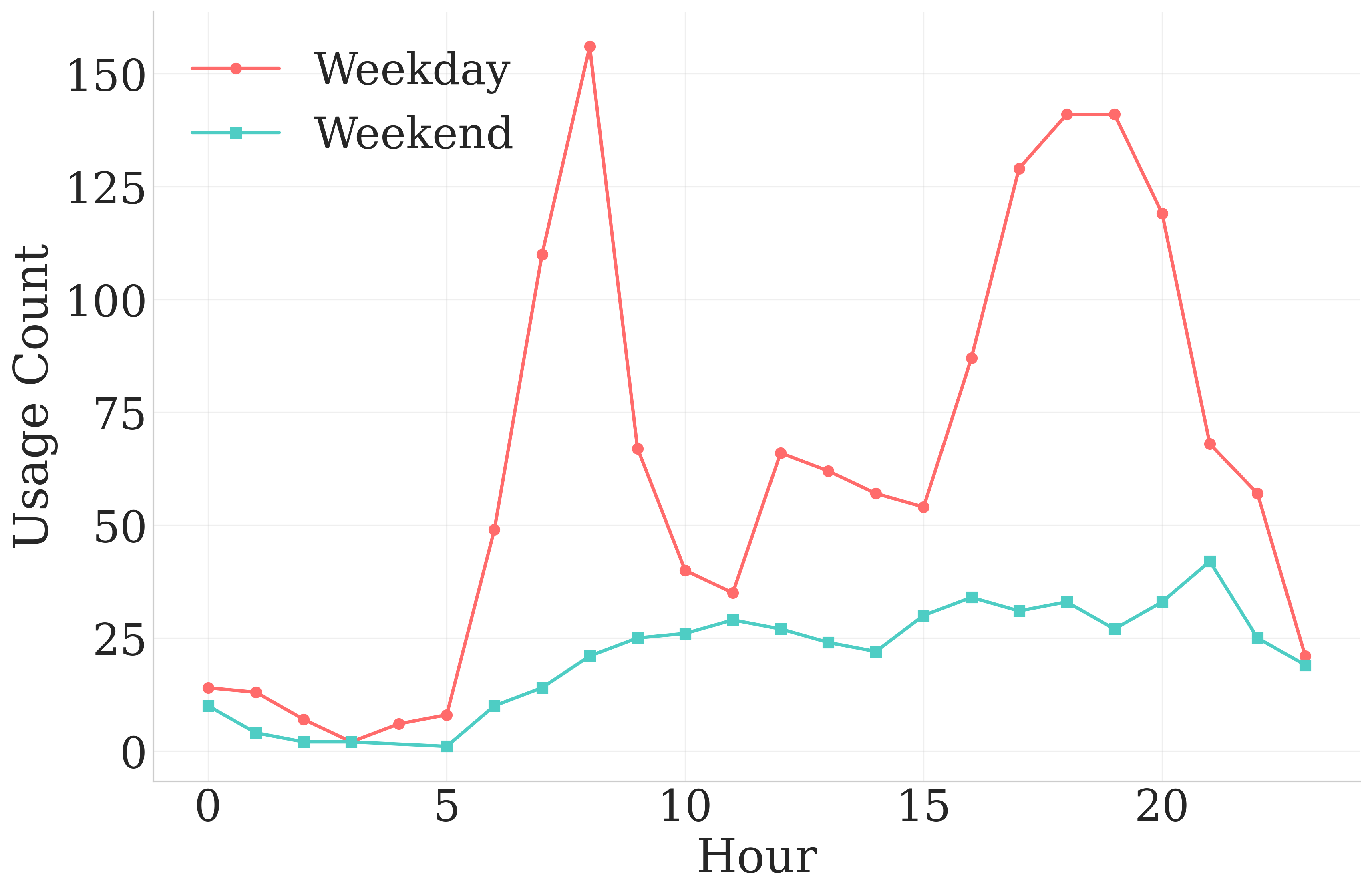}
    \caption{Functional differentiation of bike-sharing usage. The chart compares the normalized hourly trip distribution on weekdays versus weekends. It reveals two distinct patterns: a sharp, bimodal commute pattern (Utilitarian Tool) and a broader, flatter leisure pattern (Recreational Toy).}
    \label{fig:two_patterns}
\end{figure}

Figure~\ref{fig:two_patterns} reveals a stark contrast in these temporal rhythms, confirming a clear functional differentiation. The weekday usage pattern, which accounts for 62.4\% of all trips in our dataset, is defined by a classic bimodal distribution. Activity sharply peaks during the morning (08:00-09:00) and evening (18:00-19:00) commute hours. The peak-to-trough ratio—the ratio of activity during the busiest hour to the quietest hour (04:00-05:00)—is a high 3.2-to-1. This pronounced, structured pattern is the signature of bike-sharing as a \textbf{utilitarian tool}, deeply integrated into the rigid schedule of the work week, primarily for first-and-last-mile connections to public transit.

The weekend pattern tells a completely different story. The sharp commute peaks dissolve, replaced by a broader, flatter unimodal curve. Activity gradually rises throughout the day, reaching a more prominent and sustained peak in the mid-afternoon (14:00-16:00). The peak-to-trough ratio drops significantly to 2.1-to-1, reflecting a more spontaneous, flexible, and less time-sensitive usage style. This is the signature of bike-sharing as a \textbf{recreational toy}, used for leisure, social visits, and exploring the city at one's own pace.

Crucially, when we segment this temporal data by the economic level of the trip's origin, the functional dichotomy becomes even more pronounced. We found that weekend usage is not evenly distributed. Trips originating in high-economic areas are \textbf{45\% more likely} to occur on a weekend compared to those originating in low-economic areas. Conversely, the rigid morning commute peak (specifically at 07:30 AM) is most pronounced in lower-economic zones. This quantitative evidence supports our hypothesis: the data reveals two distinct modes of use. For many residents in lower-income areas, the shared bike is primarily a functional tool for the weekday commute. For residents in more affluent areas, it serves as both a tool and a toy—a flexible mobility option that enhances their lifestyle, particularly during discretionary leisure time.

\subsection{The Inverted U-Curve: Identifying the Middle-Class Core}
\begin{figure}[t]
    \centering
    \includegraphics[width=\columnwidth]{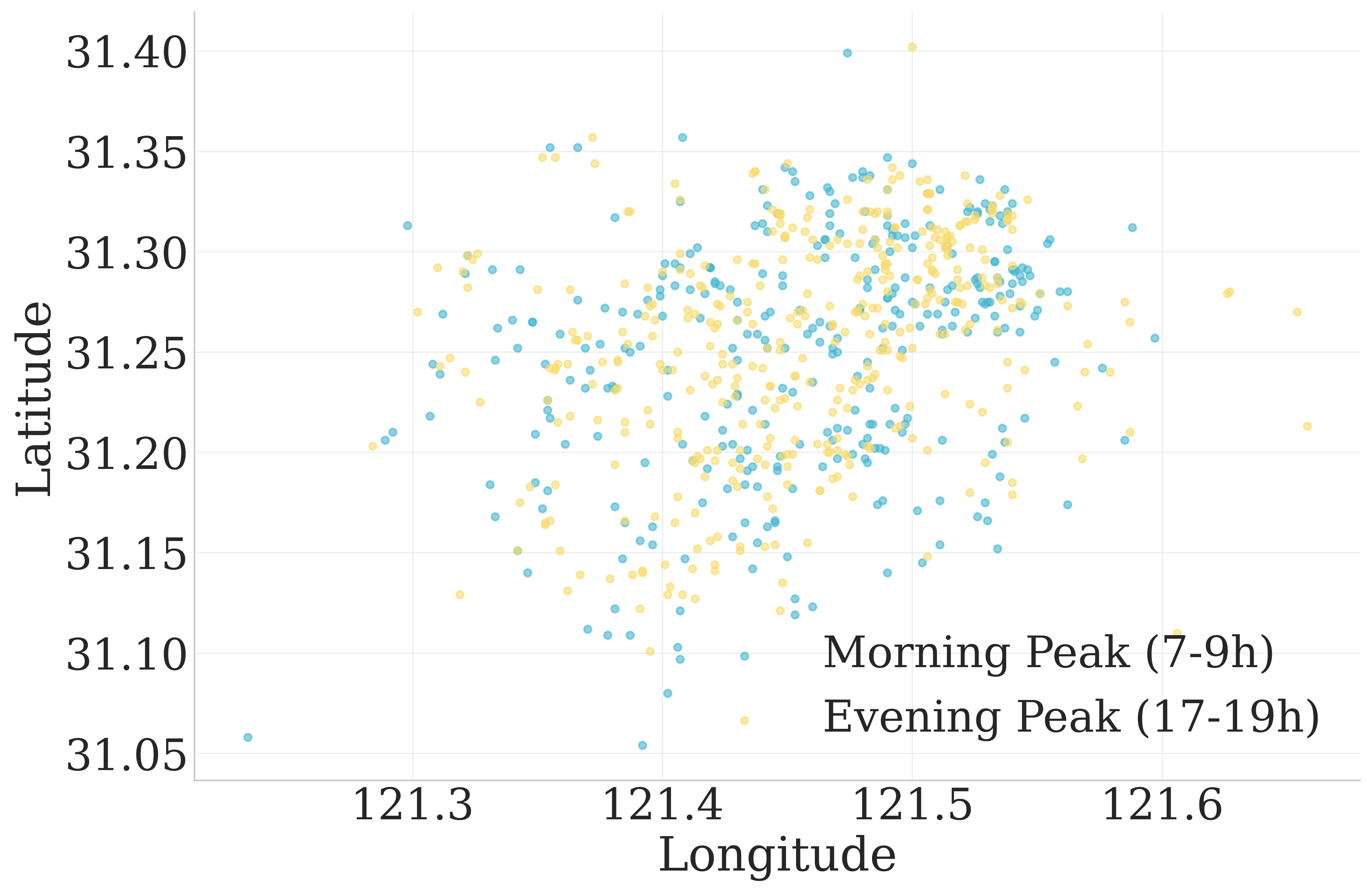}
    \caption{Visualization of the ``tidal commute'' pattern. The map illustrates the dominant directional flows of trips during morning (07:00-09:00) and evening (17:00-19:00) peak hours, revealing the structured daily rhythm of urban mobility.}
    \label{fig:tidal_commute}
\end{figure}

Our final narrative examines the collective rhythm of the city to reveal a nuanced relationship between economic status and bike-sharing adoption. While individual behaviors differ, their aggregation creates a city-wide ``tidal commute''—a structured, daily pulse of movement between residential and commercial zones. Our analysis of trip directionality during peak hours, visualized in Figure~\ref{fig:tidal_commute}, confirms this phenomenon. A particularly interesting finding from the Shanghai data is the significant outward flow (68\% of morning peak trips) from the city center, likely reflecting the unique urban structure where central districts function as both high-end residential zones and employment nexuses. This complex flow underscores the deep integration of bike-sharing into the city's multi-modal transit ecosystem.

\begin{figure}[t]
    \centering
    \includegraphics[width=\columnwidth]{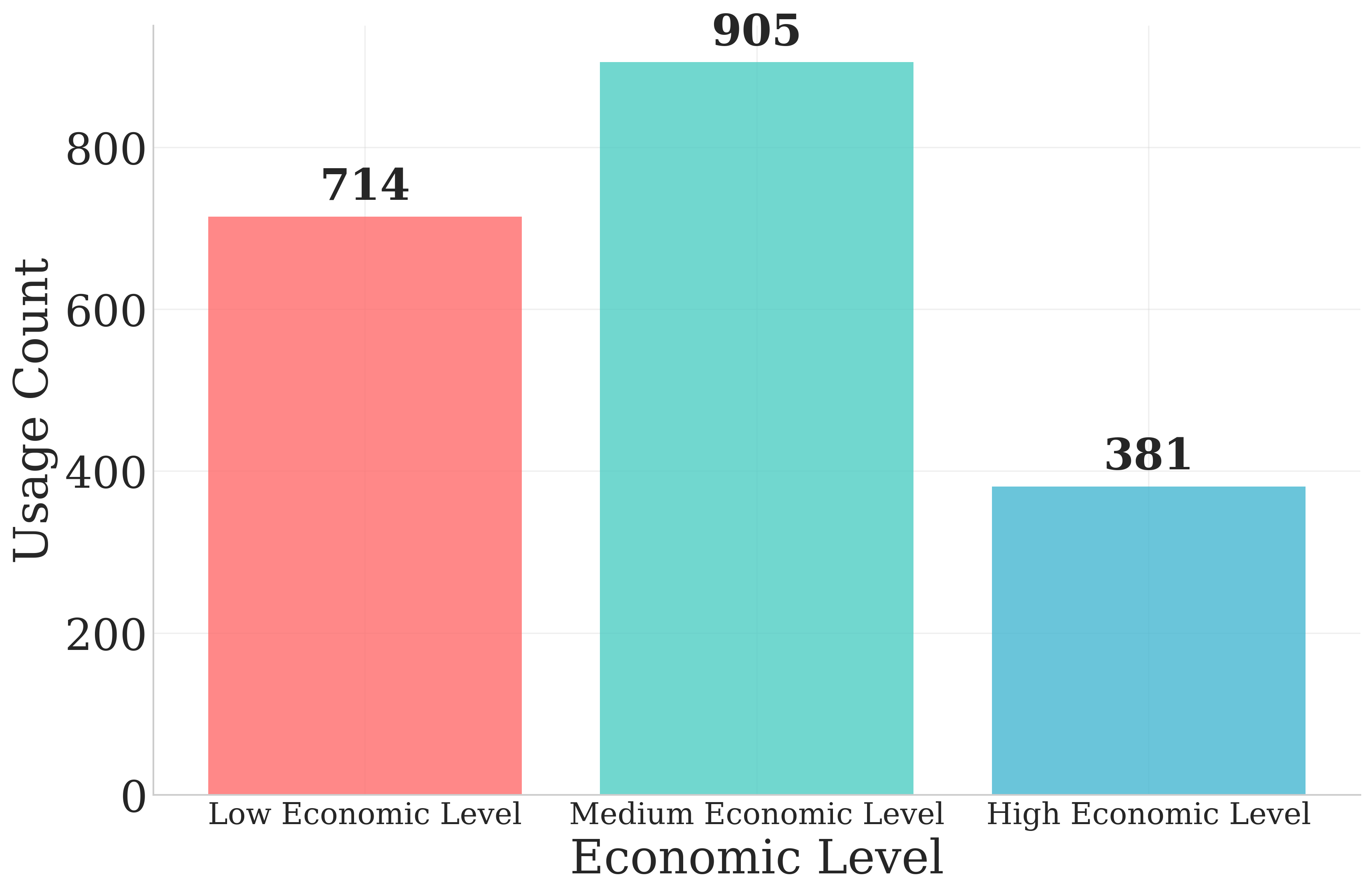}
    \caption{The relationship between economic level and bike-sharing usage. The bar chart reveals an inverted U-shaped pattern, with the medium-economic level group showing the highest frequency of use. Annotations provide key usage statistics for each tier.}
    \label{fig:economic_patterns}
\end{figure}

Perhaps the most surprising insight, however, emerges when we directly correlate economic levels with usage frequency. Contrary to a simple linear assumption that wealth directly translates to higher usage, the data reveals a more complex, inverted U-shaped relationship. To analyze this, we segmented the trip origins into three economic tiers based on housing price: Low (bottom 33\%), Medium (middle 33\%), and High (top 33\%). The results are detailed in Figure~\ref{fig:economic_patterns}.

The highest frequency of bike-sharing usage is found not in the wealthiest or poorest areas, but in the \textbf{medium-economic level} group, which accounts for 39.5\% of all trips. This group represents the city's broad middle class. In contrast, the high-economic group, despite having the shortest and most efficient rides (average duration 16.9 minutes), accounts for a smaller share of total trips (29.4\%). The low-economic group accounts for a comparable share (31.2\% of trips), but these trips are characteristically different: they are the longest on average in both distance (3.2 km) and duration (22.1 minutes), likely reflecting longer and more arduous commutes.

This inverted U-shaped pattern suggests a nuanced socio-economic dynamic. The high-income group, while benefiting from the system's convenience for short trips, likely has a wider array of private transport options (e.g., taxis, private cars), reducing their overall dependency on bike-sharing. The low-income group, while heavily reliant on the system for essential travel, may be constrained by lower service availability in their residential areas or by commute distances that are at the upper limit of what is feasible for biking. It is the middle-income group that appears to occupy the ``sweet spot''—they possess a strong need for efficient and affordable urban mobility, reside in areas with good BSS coverage, and have fully integrated bike-sharing into their daily travel routines. This finding refines our understanding of mobility inequality: it is not a simple rich-poor divide, but a complex spectrum where the urban middle class emerges as the core user base of the bike-sharing system.

\section{Conclusion}

This paper investigated the intricate relationship between shared mobility and urban socio-economic inequality. Confronting the persistent challenge of fine-grained data scarcity, we proposed and successfully implemented a novel framework that leverages a Large Language Model (LLM) as a knowledge engine to enrich bike-sharing data with localized housing price information. This scraper-free, scalable, and reproducible methodology offers a new paradigm for urban sensing, providing a powerful tool for fusing disparate data sources in urban computing.

Our empirical study of Shanghai's bike-sharing system, analyzed through an interpretable machine learning model, yielded a primary and unequivocal conclusion: the socio-economic status of a neighborhood is the single most powerful determinant of bike-sharing usage patterns. This conclusion is supported by three specific, data-driven findings. First, we identified a significant spatial coupling between BSS usage and neighborhood affluence, a phenomenon we term the \textit{club effect}, where mobility resources are concentrated in the city's high-priced core. Second, we uncovered a distinct functional dichotomy in user behavior: residents in lower-income areas predominantly use bike-sharing as a utilitarian tool for their work commute, whereas residents in higher-income areas exhibit more flexible usage patterns that include leisure and recreation. Third, our analysis revealed a nuanced, inverted U-shaped adoption curve, challenging a simple linear view of inequality and identifying the urban middle class as the system's core user base.

Taken together, these findings demonstrate that shared mobility systems, while offering significant benefits, are not neutral platforms. They are deeply embedded within, and reflective of, existing urban social and economic structures. The patterns of inequality we observed are not merely academic curiosities; they have real-world implications for transportation equity and access to opportunity. By providing a data-driven, quantitative lens on these issues, this research offers valuable insights for policymakers, urban planners, and system operators. The methodology and findings presented here can inform more targeted and equitable strategies for infrastructure deployment, service pricing, and rebalancing operations, ultimately contributing to the design of more inclusive urban mobility systems for the future.

This study also opens several avenues for future research. The LLM-based enrichment technique could be applied to more recent datasets to explore how these socio-economic patterns have evolved over time, particularly post-pandemic. Furthermore, applying this framework to other cities with different urban forms and social contexts would be a valuable step in testing the generalizability of our findings.

\section*{Data and Code Availability}
\renewcommand{\thefootnote}{\fnsymbol{footnote}}
The dataset and the Python code used for the analysis and visualization in this paper are publicly available in our GitHub repository\footnote{\url{https://github.com/limengyang1992/bike}}. The repository includes the raw data sample, the feature engineering scripts, the machine learning model implementation, and the notebooks required to reproduce all figures, tables, and results presented in this study. We believe in transparent and reproducible research and encourage the community to use and build upon our work.
\renewcommand{\thefootnote}{\arabic{footnote}}

\bibliographystyle{IEEEtran}
\bibliography{reference}



\end{document}